# SpIRIT Mission: In-Orbit Results and Technology Demonstrations

Michele Trenti, Miguel Ortiz del Castillo, Robert Mearns, Jack McRobbie, Clint Therakam, Airlie Chapman,
Andrew Woods, Jonathan Morgan, Simon Barraclough, Ivan Rodriguez Mallo
The University of Melbourne, Melbourne Space Laboratory,
Parkville, Victoria, 3010, Melbourne, Australia
michele.trenti@unimelb.edu.au

Giulia Baroni, Fabrizio Fiore
INAF/OATS, Via Giambattista Tiepolo 11, Trieste, Italy

Yuri Evangelista,
INAF/IAPS, Via del Fosso del Cavaliere 100, Roma, Italy

Riccardo Campana
INAF/OAS, Via Piero Gobetti 93, Bologna, Italy

Alejandro Guzman, Paul Hedderman
Institut für Astronomie und Astrophysik, Eberhard-Karls-Universität Tübingen, Germany

**ABSTRACT**

The *Space Industry Responsive Intelligent Thermal* (SpIRIT) 6U CubeSat is a mission led by The University of Melbourne in cooperation with the Italian Space Agency. SpIRIT received support from the Australian Space Agency and includes contributions from Australian space industry and international research organizations. Developed over the last four years and launched in a 510 km Polar Sun Synchronous Orbit in December 2023, SpIRIT carries multiple subsystems for scientific and technology demonstration. The main scientific payload is the HERMES instrument, a gamma and X-ray detector for detection of high-energy astrophysics transients (primarily Gamma Ray Bursts), and for studies of their variability at scales below 1 ms. The satellite includes a novel thermal management system for its class, based on a Stirling-cycle cooler and deployable thermal radiator, designed to cool the HERMES instrument to reduce instrumental background noise. A low-latency communication subsystem based on a sat-phone network is supporting the rapid transmission of time-critical data from HERMES to mission control and can receive spacecraft telecommands as well. SpIRIT is also equipped with a set of RGB and thermal IR cameras, connected to an on-board image processing unit with artificial intelligence capabilities for autonomous feature recognition. To effectively manage all the electrical, electronics and software interfaces between different subsystems and mission stakeholders, the University of Melbourne developed an instrument control unit (PMS) which operates all payloads. PMS also provides backup uninterruptible power to the HERMES instrument through a supercapacitor-based UPS for safe instrument shutdown in case of platform power interruptions. This paper first presents a brief mission and payload overview, and then focuses on early in-orbit results, along with lessons learned throughout the mission development and operations. Industry-developed subsystems on SpIRIT, which include the spacecraft platform, the attitude determination and control system, and a solid-fuel electric propulsion unit, are not covered in this paper. This work not only sheds light on the novelty of some of the on-board technologies onboard and on their potential impact to enable greater utilization of CubeSats for scientific missions, but also offers insights into the practical challenges and accomplishments related to developing and operating a multi-organization CubeSat with a complex array of instruments and systems.

**INTRODUCTION**

Nanosatellites and CubeSats for remote sensing experienced rapid growth in recent years, both in number of launches, now exceeding 2000 [1], and in capabilities. These platforms have moved beyond the initial applications as educational tools and have found many user-cases for scientific and commercial applications for Earth observations, communication, as well as planetary science and astrophysics [2]. A notable example of high scientific return (combined with educational goals) is the Colorado Student Space Weather Experiment, a 3U CubeSat launched in 2012 that is associated with 29 publications (conference proceedings and peer-reviewed journal articles) [3], including two in a journal with very high impact factor [4] [5]. Narrowing down consideration to remote sensing, CubeSats generally find their best applications when targeting unique (niche) applications in some parts of the electromagnetic spectrum (e.g. UV or gamma/X-ray astronomical observations that are impossible from the ground), and/or leverage applications that benefit from a (low-cost) constellation, such as guarantee all-sky coverage or rapid revisit [6].

Despite these achievements, three current challenges contribute to limiting use of CubeSats and nanosatellites: advanced thermal control (including operations of payloads that require cryogenic temperatures), edge (pre)processing of data produced by on-board sensors in excess of the downlink capability, and low-latency communications. These elements are interconnected and are typically all required for many advanced applications. For example, an order of 100 GB of imaging data could be generated each day on board by a continuous non-overlapping scanning strategy by considering the user case of an infrared remote sensing nanosatellite equipped with a moderate ground resolution imager (e.g. 10m/pixel), and a low noise 2048x2048 pixel sensor operating in the 1-5 µm wavelength range. Such concept of operations would not only require cooling of the infrared sensor to below 100K but also concurrent precision thermal control of the optical telescope assembly to limit foreground noise and focal length distortion. At the same time there is the need for efficient dissipation of the heat generated by the on-board subsystems to acquire, preprocess, store and downlink the data. All these elements are challenging in volume and surface-area constrained configurations. Furthermore, the satellite concept considered in this example would be greatly enhanced by the ability to respond rapidly to telecommanding from mission control (e.g., for on-demand repointing), and by the possibility of transmitting time-critical alerts to mission control.

All these elements are common on larger satellites. In fact, thermal control is facilitated by large radiating areas and there is sufficient volume for insulation and a dewar containing coolants such as liquid helium. Also, the power and mass budgets enable the use of dedicated satellite-to-satellite communication networks such as the NASA Tracking and Data Relay Satellite System. In contrast, CubeSats generally have very limited thermal control, with subsystems exhibiting large temperature fluctuations as the spacecraft goes in and out of eclipse, on-board data processing is restricted by computing resources, and communication with mission control is only through a few scheduled contact windows each day and with a link budget below an order of 1 GB/day even when using high throughput radios such as S-band [7].

Progress to advance the state of the art is ongoing. For example, the IRIDIUM and Globalstar satellite communication networks have been demonstrated as potentially viable solutions for low latency telecommand and telemetry retrieval [8]. Artificial intelligence and on-board (edge) computing on CubeSats has recently taken advantage of GPU-powered subsystems [9] and an actively cooled hyperspectral infrared imager has been launched on the 6U HyTi CubeSat in 2024 [10].

To address all these three aspects in a single satellite with the ambition to serve as a technology demonstrator for a future infrared remote sensing satellite, the 6U SpIRIT mission - led by the University of Melbourne (Australia) - was proposed in 2019, with funding awarded by the Australian Space Agency in 2020 and launch in December 2023. SpIRIT aims to derisk thermal, communication and on-board data-processing payloads, while at the same time supporting effectively operations of a compact gamma and X-ray remote sensing instrument for high-energy astrophysics, which has been provided through international cooperation with the HERMES Pathfinder consortium [11] and the Italian Space Agency. SpIRIT is among the first 6U Australian CubeSats, and the first space mission to fly that received funding from the Australian Space Agency since its establishment in 2018. As such, one element of the project is to promote the growth of the Australian space sector to demonstrate its capability to collaborate effectively with experienced teams based in Europe. Therefore, a made-in-Australia preference has been adopted for all spacecraft and mission elements funded in Australia, even though this increased risk because of very limited space heritage.

This paper presents a brief mission overview in Section 2, as well of the hardware developed by the University of Melbourne (Section 3) and of the HERMES instrument (Section 4). Section 5 includes highlights from the first six months of in-orbit operations, dedicated to the commissioning of the satellite, and Section 6 reflects on some lessons learned. Section 7 briefly summarizes and concludes.

**SPIRIT MISSION OVERVIEW**

The SpIRIT CubeSat mission is led by the University of Melbourne (Principal Investigator Michele Trenti) in cooperation with the Italian Space Agency and its associated entities. The primary aims are (1) to advance high-energy astrophysics from space by demonstrating in-orbit functionality of a novel gamma/X-ray scientific instrument (the HERMES payload) and (2) to actively control its temperature via a Stirling-cycle cryocooler; (3) to grow the Australian space industry's collaborative capability to develop and operate a complex, multi-stakeholders nanosatellite, building knowledge, technology and workforce; and (4) to inspire the (Australian) public. Funding for the project in Australia has been provided by the Australian Space Agency through two grant programs, the International Space Investment - Expand Capability scheme (to cover activities from preliminary design to delivery of the spacecraft to the Launch Service Provider) and the Moon to Mars - Demonstrator Mission scheme (to cover in-orbit operations). The mission was proposed in December 2019, funded in June 2020 and the spacecraft has been launched in a 510 km Polar Sun Synchronous orbit on December 1st, 2023.

The SpIRIT spacecraft is a 6U-XL (11.8 kg) CubeSat based on a commercial spacecraft platform with deployable solar panels, provided by Inovor Technologies (a company located in Adelaide, Australia). The spacecraft platform had no flight heritage prior to SpIRIT's launch. Approximately four litres of volume are allocated to payloads, with about three litres occupied by University of Melbourne and Italian Space Agency hardware, and one litre by a novel high-efficiency ion thruster developed by Neumann Space (a company located in Adelaide, Australia). The payload stack includes a commercial-off-the-shelf S-band transceiver provided by the Italian Space Agency. Figs 1-2 show the spacecraft being loaded into the flight pod and a schematic CAD that illustrates the spacecraft and its subsystems, respectively.

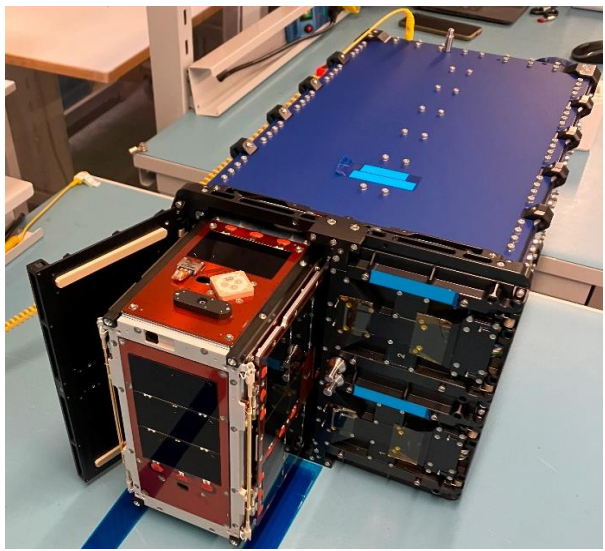

**Figure.1. The SpIRIT CubeSat during loading into the flight pod, the final integration step for the 510km Sun Synchronous Polar Orbit launch on December 1st, 2023.**

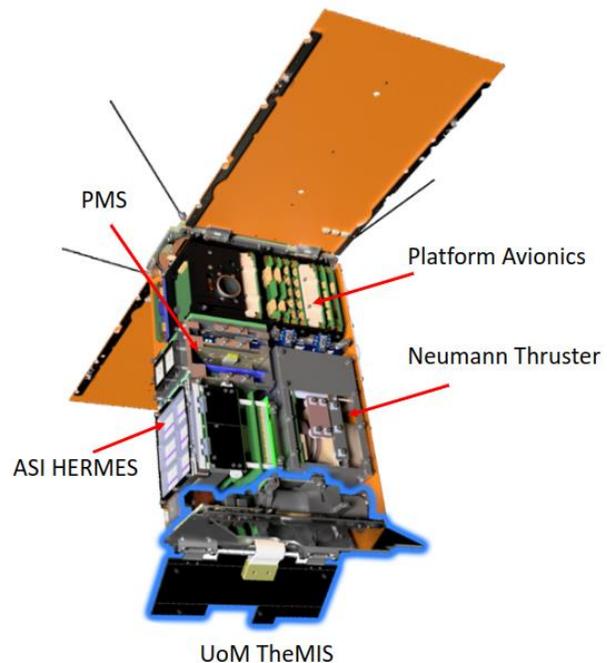

**Figure 2. Illustrative SpIRIT subsystem block diagram. Note the deployable radiators for the thermal management system (TheMIS).**

The main instrument payload is an X-ray and gamma ray remote-sensing detector (HERMES), provided by the Italian Space Agency (ASI) and developed by the HERMES SP consortium (Principal Investigator Fabrizio Fiore). The University of Melbourne's Thermal Management Integrated System (TheMIS) is designed to actively cool HERMES and its electronic controller, dissipating the heat through high-performance radiators fabricated with advanced materials.

The mission ground segment is coordinated by a dedicated control centre at the University of Melbourne, which is responsible for defining operation plans, scheduling of activities, and coordination of telemetry analysis. The primary ground stations are based in Australia and specifically Peterborough (South Australia; UHF antenna for downlink and uplink, managed by Nova Systems) and Katherine (Northern Territories; S-band downlink). A secondary S-band ground station is expected to become available in the future as an in-kind contribution by ASI.

The SpIRIT spacecraft and payloads have been designed for a notional 24-month lifetime. Given the actual altitude of deployment and the intense solar activity, current extrapolation of the orbit decay predicts an end of mission in October 2025, just short of the target mission duration.

This paper covers only elements of the mission related to the University of Melbourne and Italian Space Agency/HERMES team payloads and data retrieved from the spacecraft associated to those payloads. All elements related to Australian industry contributions to SpIRIT beyond publicly available information are omitted.

**UNIVERSITY OF MELBOURNE PAYLOADS**

In addition to its role as mission lead and system integrator, the University of Melbourne developed four main payloads on board SpIRIT: the Instrument Control Unit (PMS), the payload thermal management system (TheMIS), a module for edge computing and multi-camera control (LORIS), and a low-latency communication subsystem (Mercury).

**PMS.** The SpIRIT Instrument Control Unit - named Payload Management System (Figure 3), or PMS in short, consists of two primary components, an onboard computing unit (OBC) and an electrical power system (EPS) to interface the platform with multiple payloads. The overall design aims to allow flexible customization, with the idea that revisions will naturally accommodate alternate requirements for future projects.

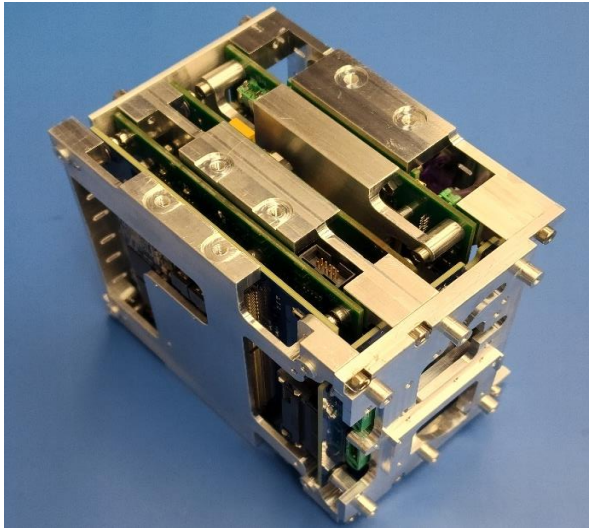

**Figure 3: SpIRIT Instrument Control Unit (PMS) Flight Model stack, inclusive of the edge-computing and multi-camera controller module for LORIS.**

The OBC unit is based on a System on Chip (FPGA & Cortex-M3) with heritage in low-Earth orbit. The system includes a variety of configurable components, and can support multiple UART, SPI, and CAN interfaces, limited only by pin availability and hardware support on the processor itself. PMS also has storage for payload data in a non-volatile way using NAND flash memory.

On SpIRIT, the payload features a fully redundant CAN bus connection to the platform. The system has been designed to be upgradeable to a radiation-tolerant FPGA and processor.

The EPS is controlled by the OBC and has a collection of regulator monitors and switches to support payload power requirements. It uses two separate power supplies in a fully redundant fashion, allowing for safe operation if there is an unexpected shutdown of one of them. The EPS also includes a supercapacitor as an emergency backup power system that is configured to support safe payload shutdown in the event of a temporary power failure from the platform. In the configuration for SpIRIT, it provides about one minute of emergency power (~5W) for the PMS itself and for the HERMES instrument.

The PMS manages payloads and meets each of their requirements through a scheduler developed in C/C++ with a FreeRTOS operating system. This philosophy aims to simplify software interfaces to payloads while retaining the ability to implement more complex and automated error handling and monitoring for payloads that require such elements. Thanks to the PMS, such operations do not need to be performed by customization of the platform itself and can be handled as a separate module instead, reducing the need to require customization of commercially available platform solutions.

**TheMIS.** TheMIS is a thermal management module designed to provide active cooling and precision temperature control for hosted instrumentation. The TheMIS requirements have been set considering applications for infrared remote sensing that require cryogenic temperatures down to T~80 K at the cooler's cold tip, although operational parameters such as temperature can vary depending on specific mission requirements and are fully tunable. A detailed description of the subsystem is presented in a companion paper [12], here, we briefly summarize the key elements of the subsystem. TheMIS comprises of passive and active thermal elements. The key active element is a Stirling cycle cryocooler driven and controlled by a custom-designed electronic and software module. Passive thermal control is aided by two deployable thermal radiators, each approximately of 2x200 $cm^2$ area, which increase by about 5 times the available area for radiative cooling of the cryocooler body. The deployment of the radiators is controlled by a small onboard computer that receives command from the PMS to actuate a commercial off-the-shelf memory-shape alloy pin puller. The pin puller releases a custom designed and highly adjustable latching mechanism.

The cryocooler provides actuation when actively driven, with the closed-loop control algorithm based on input from 8 thermal sensors located on critical locations across the cryocooler assembly and the HERMES instrument.

**LORIS**. Loris is an imaging and edge processing payload that features six visible light cameras, three thermal IR sensors and an Nvidia Jetson Nano general-purpose Graphis Processing Unit (gGPU). The imagers are controlled by separate multiplexing electronics that allow for up to 9 sensors with control via an I2C interface and data over SPI and CSI streams.

The 9 cameras (three IR Lepton 3.5 and six visible IMX219) are mounted to provide fields of view that can validate spacecraft deployable elements. All but one of the cameras are deployed on small deployable slides located on the spacecraft sides and stowed behind the stowed solar panels. A single deployment mechanism is designed to extend the slides slightly outside the spacecraft structure to enable full imaging of the radiator panels by the three thermal IR cameras, allowing for characterization of their performance on orbit. Prior to deployment of the sliders, the cameras field of view is partially obstructed by the spacecraft chassis. The visible cameras have a ground resolution of about 200 m per pixel for a 500 km orbit. One visible-light camera is mounted on a deployable arm that is automatically released with the deployable radiators. Once deployed, this camera will be able to carry out visual inspection of the Neumann Space thruster.

The Jetson Nano gGPU was selected for demonstration of edge-processing algorithms. Loris hosts a model for detection of clouds and autonomous labelling using the visible-light images, with on-orbit fine-tuning possible. Furthermore, the Jetson Nano is used for progressive image compression through JPEG-XL [13]. A detailed description of the design and capabilities of the Loris payload is presented in [14].

**Mercury**. The Mercury payload (Figure 4) provides an experimental testbed for the characterization of existing satellite communication infrastructure (SATCOM), to better understand how these SATCOM networks can be utilized by nanosatellites to supplement traditional communication links and fill gaps in high-timeliness but low data-rate communication capabilities. A reduced time to message transmission and receipt will enable the increased utilization of the nanosatellite form-factor for research and commercial applications which explicitly leverage the small agile nature of the form-factor.

The major challenge to the utilization of these networks is that they are designed for surface users. Consequently, any single network has coverage gaps in low-Earth orbit [8]. Mercury carries two COTS SATCOM modems, namely an Iridium 9603 Short Burst Data (SBD) modem and a Globalstar compatible Near Space Launch (NSL) EyeStar simplex modem. Unfortunately, late in the development of the mission, Globalstar ceased supporting orbital modems. Mercury is conducting characterization experiments using its remaining operative modem to better understand the eccentricities of the Iridium network and how these will ultimately impact usage by nanosatellites in orbit.

Additionally, Mercury provides both high precision timing information for use by the HERMES instrument, and temperature monitoring of the payload stack by controlling multiple distributed sensors.

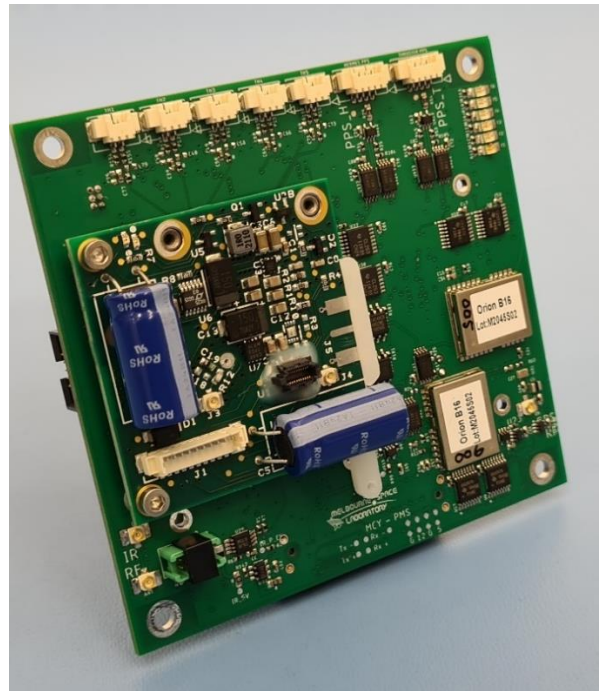

**Figure 4. Mercury duplex low-latency (Iridium) communication system Flight Model board. The board also manages redundant GPS receivers and multiple distributed temperature sensors for thermal monitoring across the SpIRIT payload stack.**

## THE HERMES INSTRUMENT

The main scientific payload on board SpIRIT is an instrument for high-energy astrophysics (Figure 5), developed by the HERMES (*High Energy Rapid Modular Ensemble of Satellites*) Technologic and Scientific Pathfinder collaboration [15]. The payload has a volume of approximately 1U and hosts a compact and innovative X-ray and gamma-ray detector designed to detect and localize high energy transients, such as Gamma Ray Bursts and the electromagnetic counterparts of Gravitational Wave events, over a wide energy range from 3 keV to 2 MeV. The instrument has a nearly "half-sky" field of view, with its effective area scaling roughly as $\cos(\theta)$, where $\theta$ is the angle between the source and the normal vector to the detector. Given the large field of view of the detector, the SpIRIT/HERMES instrument imposes modest attitude determination and control requirements on the mission, with attitude knowledge and stability both within 3 degrees (one axis, one sigma). The default HERMES operating mode assumes pointing of the instrument to the sky, avoiding both the Sun and the Earth in its field of view.

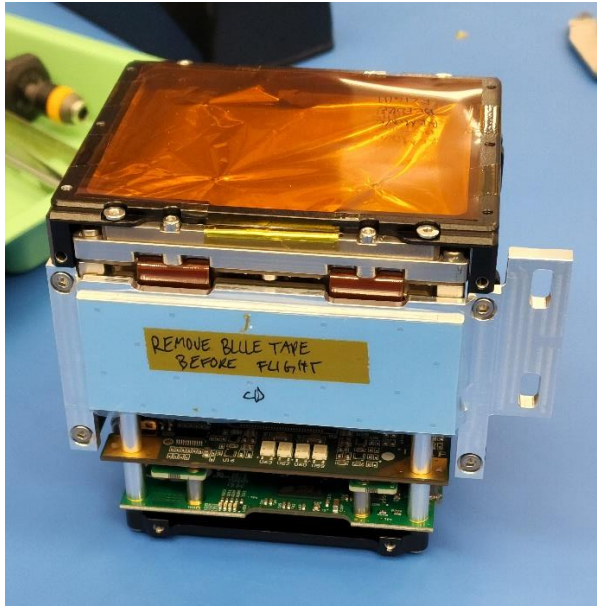

**Figure 5. HERMES instrument Flight Model, inclusive of dedicated thermal radiators (one visible in front of the image) and of a thermal blanket/optical filter in front of the detector (orange element on the top).**

The unit on SpIRIT is identical to those that will fly onboard the HERMES Technologic and Scientific Pathfinder constellation, a group of six 3U CubeSat scheduled for launch in Low Earth Orbit in the first half of 2025. Localization of sources in the sky will be possible through triangulation by precise measurement of photon arrival times once the constellation is in place [11]. The HERMES instrument design, integration and testing are described in [16] [17]. On SpIRIT, the unit is thermally connected to the cold tip of the Stirling cooler in the TheMIS subsystem, and otherwise thermally insulated from other spacecraft subsystems, with the aim to control its operating temperature and reduce temperature-dependent instrumental noise.

We note that the HERMES unit delivered for integration in the SpIRIT satellite only had three detector quadrants functional out of four, leading to a corresponding 25% reduction in the instrument collecting area. The impact on the SpIRIT mission objectives was deemed to be minimal and fully acceptable as only a modest loss on the signal to noise of observations stems from the effective area reduction. Furthermore, none of the technology and scientific demonstration objectives are impacted.

## FIRST IN-ORBIT RESULTS

The payload commissioning activities have been ongoing since launch and are providing very valuable experience to the University of Melbourne Mission Operations Center and to HERMES instrument team. SpIRIT is providing basic telemetry through an unencrypted UHF beacon every 60s, and this information, captured regularly by the amateur satellite radio community, has provided very valuable information, especially in the first few hours after launch. The University of Melbourne Mission Control Center has been in daily communication with the satellite through a UHF link, with several transmissions recorded on SatNOGS [18].

Six months post launch, all The University of Melbourne and Italian Space Agency hardware has shown no evidence of damage compared to post integration and environmental testing conditions. Functionalities are progressively being demonstrated, with several key mission objectives already achieved. The commissioning activities are taking longer than the nominal six-month period in the pre-launch schedule, though. Contributing factors to the delay in concluding commissioning include but are not limited to the lack of time in the pre-launch schedule to carry out rehearsal of payload operations under realistic conditions using flight hardware, and to the lack of a full engineering model of the spacecraft. A brief description of the commissioning status of the payload subsystems is as follows.

**PMS.** The instrument control unit has been the first payload to be commissioned following successful communications with the satellite. A basic short functional test to verify integrity of the four on-board memory chips was performed on December 5, 2023, demonstrating nominal electrical and electronic interfaces with the subsystem. Subsequently, all telecommands to control payloads connected to PMS have been tested, demonstrating full functionality of the unit. At the time of writing, PMS is fully commissioned and operates nominally, with some occasional occurrence of radiation upset events, discussed at the end of this section.

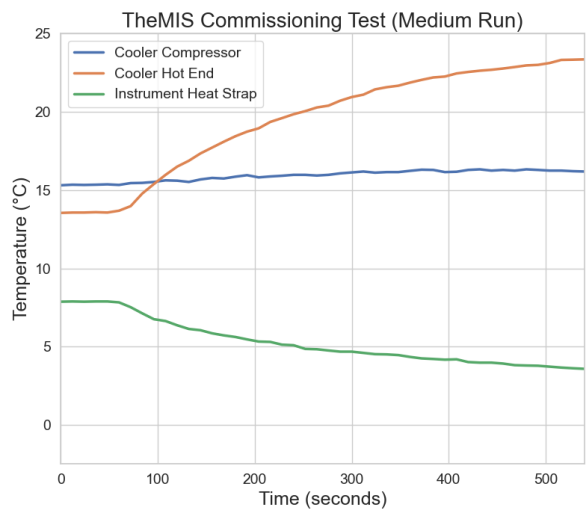

**Figure 6. TheMIS cooling test during commissioning. The decrease in the instrument heat strap temperature and the concurrent rise in the cooler hot end demonstrate the functionality of the active thermal control.**

**TheMIS**. The subsystem commissioning strategy has been based on verification of the electronics functionality first, including temperature sensing, followed by cryocooler operations. TheMIS has successfully operated the cryocooler on several occasions, employing different configurations both in an open and closed loop to demonstrate its cooling capabilities. Throughout these tests, the system has consistently shown its ability to manage thermal loads effectively, with preliminary assessment of in-orbit performance indicating consistency with pre-flight ground testing. However, it is important to note that the deployable radiators, which have a main role in the overall thermal management system, have not yet been deployed. In fact, during the current commissioning phase, payload operations have been relatively limited, and the mission has taken a risk-adverse approach to passive overcooling of the payload end of the spacecraft. Consequently, the full operational capability of TheMIS is yet to be fully verified as of 27th May 2024.

Figure 6 illustrates the results from a test lasting 500 seconds. During this operation, a reduced gain configuration was employed, resulting in a temperature drop of 5 degrees on the instrument heat strap. We note that by design the strap is connected to a large thermal mass, so the temperature decrease reflects nominal cryocooler performance under the circumstances. These initial results underscore the potential efficiency of the cryocooler even without the deployable radiators in operation. Future tests will provide more data on the system's performance, particularly once the radiators are deployed.

**LORIS**. LORIS has demonstrated successful image acquisition from all sensors and cameras to date, and several images have been retrieved through the UHF link using the JPEG-XL progressive compression algorithm [14]. Cameras on one side of the slider are currently showing uniform "dark" fields, consistent with their field of view obstructed by an undeployed solar panel. No deployable elements of LORIS have been actuated to date.

The LORIS images have been utilized to independently determine the satellite attitude through matching of the on-board data to recognizable coastlines and geographic features (Figure 7).

The on-board cloud detection and image classification capabilities of LORIS have been demonstrated on a small number of images. Further experimentation with the artificial intelligence capabilities of LORIS, including on-orbit fine-tuning of the machine learning model, is ongoing.

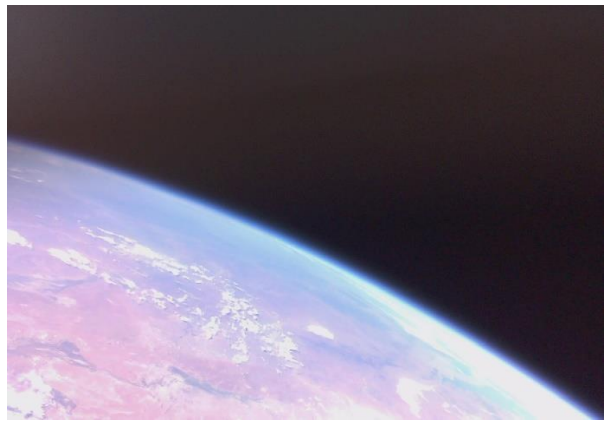

**Figure 7. LORIS IMX219 limb image captured over Australia, highlighting key geographical features and atmospheric conditions.**

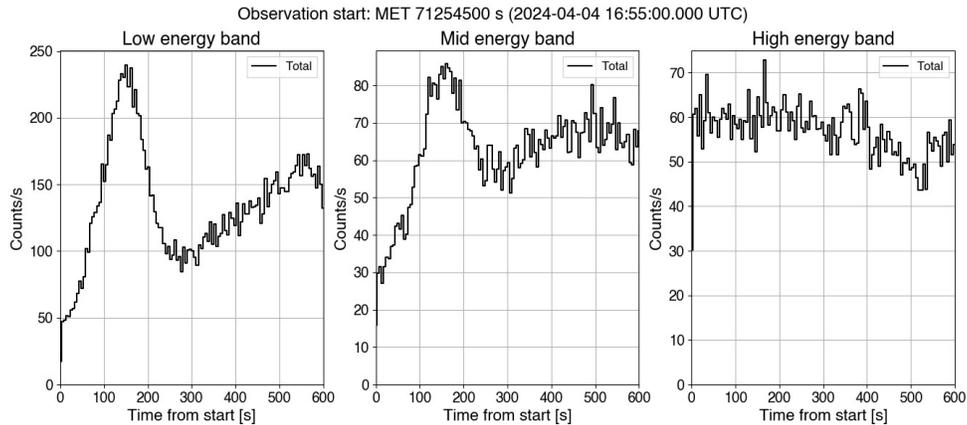

**Figure 8 Sample HERMES observation in three energy bands. During the 600s of observations the sky background counts are slowly varying. This is interpreted as partial and time-varying occultation of the instrument field of view by Earth.**

**Mercury**. Nominal behavior of all electrical and COTS systems has been demonstrated, with spacecraft temperature monitoring being utilized during Launch and Early Orbit Phase (LEOP) activities throughout December 2023. A GPS time and position lock was first acquired on January 22$^{nd}$, 2024, with intermittent success since that time in maintaining a position lock throughout spacecraft operations. Analysis and amelioration efforts are ongoing.

Mercury has successfully received ring channel signals from the Iridium network, but at the time of writing has been unable to complete connection attempts, likely due to position information from the GPS being stale at the time of submission to the Iridium network. The payload is still undergoing commissioning.

**HERMES**. Given that the instrument on SpIRIT is the first of the constellation to reach orbit, the HERMES payload commissioning has been focused on technology verification and establishment of flight heritage under all operating conditions, with the specific aim to characterize in-orbit functionality and performance. The commissioning process involved a sequential iterative procedure to verify all its operating modes. This included the acquisition of housekeeping data, monitoring various parameters such as voltages, currents, and temperatures, and the analysis of telemetered data retrieved to verify the payload health. Each operating mode, from Standby to Observation, was meticulously tested to ensure system stability and safety before and during scientific data acquisition.

After completing initial checks which confirmed that the hardware health status was consistent with the conditions recorded during pre- and post-integration ground testing, the detectors were activated, achieving "first light", on January 16, 2024. For this acquisition the instrument was operated in a mode similar to a Geiger counter. Subsequently, on February 17, 2024, the payload entered Calibration mode, with the recording of the full spectral and temporal data associated with each photon detection event, where the refinement of operational calibration parameters played a key role in improving data quality by mitigating noise and optimizing the extraction of useful information. On March 27, 2024, HERMES successfully entered nominal Observation mode, recording scientific data (light curves, see Figure 8). These first scientific data have demonstrated the operational capabilities of the instrument, with background noise levels consistent with pre-flight estimates and the time variability of the counts consistent with partial drift of Earth into the field of view of the instrument [19]. At the time of writing the HERMES instrument is operating nominally. As the mission progresses, continuous data collection and analysis will further refine the payload calibration and optimize its operational efficiency, aiming to achieve detection of astrophysical transients (Gamma Ray Bursts) and of known bright and time-variable gamma/x-ray sources such as the Crab Nebula.

**S-band**. The S-band unit has been verified as functional, and a test transmission of a Fibonacci sequence has been successfully received by the SpIRIT S-band ground station and decoded on 24 May 2024. Commissioning of this system, which requires more stringent satellite pointing requirements compared to a typical UHF space to ground communication, is ongoing.

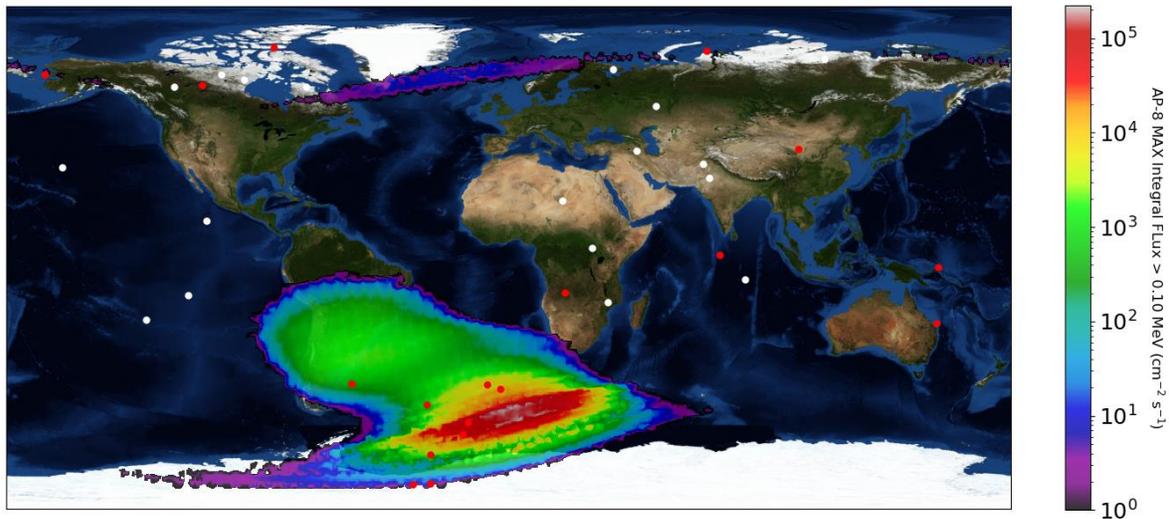

**Figure 9. Payload Anomalies plotted on an Earth map based on time of event. Red dots indicate anomalies that occurred during typical space weather, while white dots are temporally associated to solar storm. Overlayed is the proton flux heat map for energies above 0.1 MeV from the Van Allen Belt.**

**Radiation upset events**. SpIRIT has been in orbit for 181 days as of the 27th of May 2024. To characterize the PMS and payload stack in-flight performance, upset events that created a "timeout", or "bad state" errors have been recorded throughout the mission. The PMS has also been intentionally kept powered on for extensive time during periods of enhanced geomagnetic activity to stress-test its radiation resilience. Analysis was performed to identify those most likely to be associated to radiation and/or charged particle interactions, rather than attributable to flight software bugs arising from (rare) edge cases, which were identified through analysis and comparison with telemetry from flight-equivalent hardware on the ground. A total of 17 events were recorded: N=5 from PMS, N=6 from LORIS, N=3 from HERMES, N=2 from Mercury and one from the S-band unit. Figure 9 shows the spacecraft location at the time of these radiation events. It is immediate to note that there is a high concentration of anomalies in the South Atlantic Anomaly region and around the poles. Another group of anomalies is temporally associated to geomagnetic storms, and it is shown in white, with a spatial distribution that appears reasonably uniform. Finally, there are a few anomalies that have a seemingly random spatial and temporal distribution. Because the specific payloads are powered only during their operations, these anomalies could be the result of radiation events in the past being discovered as a payload is turned on. A quantitative analysis of the events to determine their degree of correlation with solar activity and orbital position will be carried out in a future work.

## SOME LESSONS LEARNED

SpIRIT is the first space mission for the University of Melbourne (Melbourne Space Laboratory) and as such it represents an ongoing opportunity to gain very valuable experience. Here, we briefly summarize some lessons that the team reflected upon to date.

*Clear definition of requirements*. SpIRIT was designed with a top-down approach, starting from a more ambitious goal of a 12U CubeSat to host an actively cooled infrared space telescope and identifying which elements of technology development could be demonstrated using a remote sensing instrument such as HERMES, which has far less stringent thermal management and pointing requirements compared to an IR imager. The clarity in scope was very beneficial to guide design and decision making by the University of Melbourne Principal Investigator and by the team during all mission phases.

*Scope creep avoidance*. Enthusiasm for increasing design complexity is anecdotally common during mission development, and our team experienced it directly. With the aim to avoid schedule delays and cost overruns, the satellite design has been kept as simple as possible to meet the key requirements. The only exception has been the inclusion of the AI and imaging payload LORIS, after careful consideration of the cost/benefit analysis and the success in securing additional external funding to support this extra effort.

*Priorities for descoping*. The SpIRIT mission started during the COVID-19 pandemic based on a pre-pandemic proposal and funding request. Supply chain disruptions, work-from-home directives and sick leave absences challenged the team to achieve the mission requirements within the timeframe and budget constraints. As a consequence, a philosophy of defining a minimum viable product (MVP) and a priority for descoping functionality and/or testing of subsystems was adopted. This approach allowed us to successfully navigate the development challenges, albeit the in-orbit activities are seeing some impact from choices to reduce documentation and/or testing of subsystems that were not deemed part of the MVP.

*Leveraging prior expertise*. Given the experimental and ambitious nature of the mission, and the absence of an institutional knowledge on space projects at the University of Melbourne, a key element for achieving success in orbit has been the prior expertise in complex space projects by several team members in the mission office. A mix of past contributions to both nanosatellite projects and traditional large space missions has been particularly useful to prevent and/or mitigate mishaps. Our team also took advantage of the general openness to sharing of advice and lessons learned in the settings of scientific missions, and reached out to a wide network of experienced collaborators and colleagues for advice.

*Internships*. One of the most rewarding aspects of the mission has been the participation of enthusiastic and committed students, through a paid internship program that saw participation of nearly 20 undergraduate and graduate students. These internships have also been very valuable as a recruitment tool, and the current mission operation team at the University of Melbourne includes three members who started their contributions to the project as interns prior to launch.

*Engineering model and/or flight equivalent hardware availability*. It is not uncommon for CubeSats to adopt a proto-flight model philosophy, and our schedule and budget constraints drove us to this approach. However, an effort has been made to develop and utilize as much flight-equivalent hardware as possible, both for early testing pre-flight and for the ongoing telecommanding preparation and validation. Our current experience with operations highlights that analysis of telemetry and debugging of software issues is enormously facilitated by availability of representative hardware in the laboratory. The commissioning of SpIRIT subsystems for which such hardware is not available are much more time and resource intensive during operations, and likely the lack of hardware is not cost-effective over the full mission lifetime.

*Project management resourcing*. The funding scheme cost cap implied difficult choices in terms of staff roles, and the project manager role was blurred into the systems engineer role. The project delivered adequate mission documentation for its class and risk profile up to the critical design review, but as the focus shifted on manufacturing and testing, we had to drop the standard for monitoring documentation of design changes as well as ground test setup and results. Furthermore, the development activities across a geographically distributed consortium comprising of multiple partners would have in retrospect required a full-time project manager to support effectively the Principal Investigator and the systems engineer. Adverse consequences were experienced during systems integration, where some mechanical and electrical interface mismatches emerged and led to schedule slippage. Missing details in the final software interfaces and testing documentation have also impacted negatively the efficiency of payload commissioning.

*Personnel backup roles*. Implementing effective duplication of expertise (i.e. backup roles) within a small mission team in a research institution setting proved to be extremely challenging, if not impossible. This implied schedule slippage when personnel were on leave, with the COVID-19 pandemic compounding the negative impact and introducing extra risk. Our experience suggests that backup roles should be strongly considered when resourcing small missions.

*Day in life rehearsals*. The SpIRIT development had non-movable schedule constraints, determined by a combination of fixed funding amount, need to spend the funding by a set deadline and desire of industry partners to gain flight heritage in a timely way. This resulted in descoping of planned activities. One element that was not carried out prior to launch was day-in-life rehearsals of early mission operations. While we have no evidence that this has negatively impacted the ability to achieve in-orbit demonstration of functionality, the time needed to commission the subsystems has been clearly extended. Therefore, the impact of skipping these pre-flight activities should not be underestimated.

**SUMMARY AND CONCLUSIONS**

In this paper, we presented a brief overview of the 6U SpIRIT satellite, which was launched on December 1st, 2023, and is equipped with technology and scientific demonstration payloads. The SpIRIT mission, led by the University of Melbourne in collaboration with the Italian

Space Agency, aims to advance high-energy astrophysics by demonstrating the in-orbit functionality of the HERMES gamma/X-ray scientific instrument, to develop a novel thermal management system and to give flight heritage to various other systems. The satellite also aims to enhance the Australian space industry's capabilities and inspire the public.

We summarized the current payload health status and early in-orbit commissioning achievements. To date, The University of Melbourne and Italian Space Agency hardware has shown no evidence of damage post-launch, with several key mission objectives already achieved. The Instrument Control Unit (PMS) has been fully commissioned and operates nominally. The TheMIS subsystem has successfully demonstrated its cooling capabilities, although the deployable radiators are yet to be deployed. LORIS has successfully captured and transmitted images, and its edge processing capabilities have been validated, albeit on a small number of cases. The Mercury subsystem has shown nominal electrical behavior and is undergoing further commissioning. The HERMES instrument has successfully recorded its first scientific data, operating in Observation mode. With SpIRIT having passed the mark of six months in orbit, these achievements are notable, given the class of the mission, the complexity of the subsystem interfaces, and the lack of prior flight heritage.

SpIRIT has also provided numerous lessons learned. Positive key lessons include the importance of clear definition of requirements, scope creep avoidance, prioritization for descoping, leveraging prior expertise, and the value of internships in university-led small satellite projects. We also highlighted the challenges stemming from the lack of full engineering models, and the importance of fully resourcing project management activities, emphasizing the need for adequate documentation and testing. Effective personnel backup roles and pre-launch rehearsals were also identified as elements that fell short of ideal in SpIRIT. These would be critical for increasing efficiency in future projects.

Overall, we assess the mission as a success based on current achievements, and there is a positive outlook for entering sustained payload operations. We consider this remarkable given the COVID-19 pandemic disruptions and especially taking into account that SpIRIT was the first mission for the Melbourne Space Laboratory at the University of Melbourne. Irrespective of the next steps in orbit, the lessons learned and risk mitigation in hardware development have positively contributed to the Melbourne Space Laboratory's ambition to develop an actively cooled IR space telescope on a CubeSat (or small satellite), and to raise its visibility for thermal engineering space projects.

*Acknowledgments*

This research received funding from the International Space Investment - Expand Capability grant ISIEC00086 and the Moon to Mars - Demonstrator Mission grant MTMDM00034 from the Department of Industry, Science and Resources (Australian Space Agency). This research was also supported in part through the Australian National Intelligence and Security Discovery Research Grant NI0100072. MT acknowledges support in part from the Australian Research Council Centre of Excellence for All Sky Astrophysics in 3 Dimensions (ASTRO 3D), through project number CE170100013.